# Growth and texture of Spark Plasma Sintered $Al_2O_3$ ceramics: a combined analysis of X-rays and Electron Back Scatter Diffraction

D. Pravarthana[1], D. Chateigner[1,2], L. Lutterotti[3], M. Lacotte[1], S. Marinel[1], P.A. Dubos[1], I. Hervas[1], E. Hug[1], P. Salvador[4] and W. Prellier[1]

[1] Laboratoire CRISMAT, Normandie Université, CNRS UMR 6508, ENSICAEN,
6 Boulevard Maréchal Juin, 14050 Caen Cedex, France

[2] Institut Universitaire de Technologie (IUT), 6 Boulevard Maréchal Juin, 14032 Caen Cedex, France

[3] Department of Materials Engineering, University of Trento,
via Mesiano, 77 - 38050 Trento, Italy

[4] Materials Science and Engineering, Carnegie Mellon University
149 Roberts Eng Hall, Pittsburgh PA 15213

**Abstract**

Textured alumina ceramics were obtained by Spark Plasma Sintering (SPS) of undoped commercial α-$Al_2O_3$ powders. Various parameters (density, grain growth, grain size distribution) of the alumina ceramics, sintered at two typical temperatures 1400°C and 1700°C, are investigated. Quantitative textural and structural analysis, carried out using a combination of Electron Back Scattering Diffraction (EBSD) and X-ray diffraction (XRD), are represented in the form of mapping, and pole figures. The mechanical properties of these textured alumina ceramics include high elastic modulus and hardness value with high anisotropic nature, opening the door for a large range of applications.

**Introduction**

Alumina (Al$_2$O$_3$) ceramic has been widely studied for its multifunctional applications in electronics, mechanical, biomedical, chemical, optical, refractory military uses and enameling properties. [1-4] In particular, many efforts are focused on the control of the grain size, shape, porosity, texture and density in order to understand and precisely tune their mechanical properties. Despite the numerous reported works on preparing ceramic alumina, very few of them are, however, addressing a correlation between processing, microstructure, and properties.

There are various synthesis techniques available to synthesize ceramic materials and obtain the required microstructure for various applications. [5] Recently Spark Plasma Sintering (SPS) is an emerging technique that has been utilized to densify different kinds of materials. The SPS process is a pressure-assisted pulsed current sintering process in which densification is highly promoted at lower temperatures compared to conventional processes. This process usually leads to highly dense ceramics with fine control of grain structures. [6,7] SPS is indeed useful because of its versatile nature in densification in a short amount of time, avoiding the coarsening process and leading to texture. [8] It has been shown for Ca$_3$Co$_4$O$_9$ and Bi$_2$Te$_3$ that the spark plasma process could significantly induce texture along the direction of applied pressure leading to changes in the thermoelectric performance. [9-11] Thus, the texturing of alumina ceramics by controlling the SPS parameters (in terms of temperature and uniaxial pressure) is necessary to design advanced ceramics with tunable mechanical properties. The preferred crystallographic orientation (texture) of polycrystalline materials influences many properties of the bulk material at the macroscopic scale via the intrinsically anisotropic character of most of the physical properties observed at the level of a single crystal. This is particularly true for elastic properties. Hence a careful examination of the texture's effect at the polycrystalline sample scale is necessary to understand its physical impact on the orientation and the elaboration technique. Therefore, the determination of the crystallographic preferred orientation of alumina generated through SPS is required.

Several methods have been developed to quantitatively determine texture in materials, those based on diffraction techniques (such as x-ray, electron, neutron diffraction) being the most efficient. Using X-rays, intensities diffracted from a set of diffracting planes are measured while the sample is stepped through a series of orientations to complete the pole figures. Electron Back Scatter Diffraction (EBSD) technique based on electron diffraction has also proven its efficiency and is widely used for texture analysis in materials science. [12]

The aim of the present work is to precisely determine the texture of spark plasma sintered $Al_2O_3$ ceramics measured as a function of sintering temperature. Alsokey parameters like grain size, misorientation angle, and orientation distributions are investigated using combined analysis of x-rays and electron backscattering diffractions. This latter parameter is used to simulate macroscopic elastic stiffness tensors that might impede the mechanical response of the material, and to compare to experimental mechanical properties. Finally, we expect such a study to be useful for a better understanding of the growth mechanisms of corundum ceramic products which are widely used in a range of applications.

**Experimental section**

Commercially available α-$Al_2O_3$ (corundum) powder was purchased (Chempur GmbH, Germany, purity of 99.97 %). The Scanning Electron Microscopy analysis confirms the average particle size of about 8 µm diameter. The purchased powder was sintered using a spark plasma sintering apparatus (Struers Tegra Force-5). For SPS sintering nearly 5 grams were loaded into a graphite die with a 20 mm diameter with graphite paper on both sides of the graphite punch and were sintered between different temperature ranges from 1400 and 1700°C. The heating rate was kept at 90 °C/min until reaching 450 °C at under an uniaxial load of 16 MPa, and then was increased to 100 °C/min with a simultaneous increasing of the uniaxial load to 100 MPa to reach the final sintering temperature of 1400 °C and 1700 °C for the two samples, with dwell times of 20 and 10 minutes, respectively. Finally, the samples were cooled to room temperature at 90 °C/min. To remove the carbon from the surface, the samples were kept at 1200 °C for 12 hours.

The alumina pellet was then cut perpendicular to the direction of the uniaxial load. To ensure a high quality of the Kikuchi patterns on the Electron Backscatter Diffraction a highly polished surface was obtained using mechanical polishing followed by thermal etching at 1200 °C for 10 mins. The EBSD experiments were performed to determine texture on the polished surface obtained as described above with carbon tape along the edges to avoid charging. The samples were mounted in the EBSD sample holder with a 70°-tilt angle from horizontal in a SEM operated at 20 kV. Automated EBSD scans were performed with a step size of 1/10$^{th}$ of grain size; the working distance was set to 10 mm and identified the sample as the α-$Al_2O_3$ phase (space group R-3c:H, Crystallography Open Database n° 1000032).[13] The acquired Kikuchi patterns were indexed automatically by the EDAX's Orientation Imaging Microscopy (OIM$^{TM}$) software (v. 6). From this dataset, an Orientation Distribution (OD) of crystallites was calculated using discrete 5° binning input parameters and 5° Gaussian smoothing, which allowed the re-calculation of both pole figures and inverse pole figures.

Texture and structural variations were also investigated at a larger macroscopic scale by X-ray diffraction, using a 4-circle diffractometer setup equipped with a Curved Position Sensitive detector (CPS120 from INEL SA), and a monochromatized Cu Kα radiation. The overall texture strength is evaluated through the texture index which is expressed in m.r.d.$^2$ units and varies from 1 (random powder) to infinity (perfect texture or single crystal) and used to compare the texture strength of different samples exhibiting similar Orientation Distributions (OD). [14] The sample reference frame is given by the SPS direction of pressure ($P_{SPS}$), which corresponds to the centers of the pole figures (Z).

The hardness and Young's modulus of elasticity were extracted from the nanoindentation experiments performed in the mirror-like polished sample. Superficial hardness profiles were obtained with a MTS XP™ nanoindentation device using the continuous stiffness measurement mode. Nanohardness was measured using the Oliver and Pharr methodology.[15] Poisson's ratio for the test material (0.21 value is generally used for alumina) and 1141 GPa and 0.07 are the elastic modulus and Poisson's ratio of the diamond indenter, respectively. On each sample, two matrixes of indents were performed (25 indents – 5 X 5 - ; X-Y-space: 90-160 µm; indentation depth: 120 µm). The loading path consisted of five force increments with holding time of 30 s and unloading to 90 % of the latest force. The time of loading was 15 s and the maximal applied force was 5 mN. Young modulus values are measured during the 5 unloading with the slope at maximum loading and are then averaged. The extracted parameters are the mean values of 25 measurements. The density of the samples was finally measured from Micrometrics AccuPyc 1330 Gas Pycnometer.

**Results and Discussion**

Figure 1 displays the typical evolution of the relative density as a function of dwell time. A uniaxial load of 100 MPa was used to minimize coarsening and to accelerate the sintering process. From the curve, a dwell time of 20 and 10 minutes is obtained for 1400 °C and 1700 °C, respectively. Figure 1 further indicates that the densification increases linearly up to the end of dwell time much faster for the higher temperature, while at 1400°C a two-regime behavior appears before saturation. At saturation, within the spanned dwell times, the process leads to a relative density of 96 % at 1400 °C, whereas at 1700 °C the final relative density reached almost the full densification value of 99.99 %. This is consistent with an increase of atomic diffusion at grain boundaries during sintering at larger temperatures.

The EBSD analysis, performed on the 1400 °C and 1700 °C samples, is obtained within overall good qualities (see the Image Quality (IQ) maps in Figures 2a and 3a). Such IQ images are formed by mapping the IQ index obtained for each point in an EBSD scan onto a gray scale in which the darker gray shades in the image denote lower IQ values. The average image quality for the scan shown is 961 and 914 for Fig.2a and Fig.3a, respectively. Note that this is not an absolute value; it depends on the material, the technique and the parameters used to index the pattern. The overall quality maps exhibit high amounts of gray over the grains confirming a good crystalline nature and homogeneity. However, readily apparent in the IQ maps, the presence of grain boundaries with some areas appearing dark over the whole grain, points probably to rough surfaces or residual porosities.[16] There are, however, less dark areas in the IQ image of the 1700 °C sample, and grain boundaries appear weaker, at the benefit of globally lighter grains, indicating a less perturbed crystal state for higher sintering temperatures (even if the overall Kikuchi patterns do not show a clear difference). EBSD orientation maps of the sample surface perpendicular to $P_{SPS}$ illustrate the crystalline direction of each grain aligning with the pressing axis, using a color coded inverse pole figure definition (Figures 2b and 3b). A significantly larger fraction of red color grains are observed indicating that the <0001> crystal directions align with $P_{SPS}$. The {0001} and {1-210} pole figures computed from the OD (Figures 2c and 3c) exhibit similar orientations for both samples, i.e. of a prominent <0001> parallel to the $P_{SPS}$, with relatively low orientation densities at distribution maxima (around 3 m.r.d.) and a- and b-axes randomly oriented around c-axis indicating a cyclic-fiber texture.

In order to obtain a larger macroscopic view, by analyzing a larger volume in comparison to EBSD, it was coupled with X-ray diffraction analysis for overall texture determination probing around one million crystals. Strong variations of intensities in the x-ray diffraction diagrams measured for various sample orientations were observed, indicating the presence of a relatively pronounced texture. The combined analysis refinement (Figure 4) correctly reproduces the experimental diagrams, within reliability factors $R_w$ = 38.23 % and $R_{exp}$ = 25.89 %. These apparently large values of reliability factors depend on the number of experimental points, which in the present case is very large (around 2.3 millions). From these two factors, a $\chi^2$ value was evaluated to be of 1.48 and corresponds to a good refinement value.

The pole figures of the main crystallographic directions of $Al_2O_3$ (Figure 5) show a relatively strong <001> cyclic-fiber-texture for the 1400°C sample. Note that similar images were recorded (not shown) for the 1700 °C sample. A maximum of the OD of 33.7 m.r.d., corresponding to the nearly 4.5 m.r.d. maximum value of the {006} pole figure is also obtained. The <001> fiber character of the texture is confirmed by the axially symmetric distribution of the {300} pole figure (Figure 5), and is consistent with the EBSD analysis. The OD has also been refined with satisfactory reliability factors, $R_w$ = 18.11 % and $R_B$ = 18.30 %. Only one component of texture exists, counting for the whole volume, which can be compared favorably with the strongest textures observed in corundum ceramics elaborated by other techniques.[17] The refinement converges to unit-cell parameters and atomic positions close to the usual values, with no significant variation between the two samples (Table 1). In addition, the overall texture strength, maxima of OD, pole figures and inverse pole figures do not show significant variation with the sintering temperature from both EBSD and X-ray texture analysis. However, subtle differences are observed between the X-rays and EBSD pole figures in density levels. For example, the non-axial symmetry exhibited in EBSD pole figures (which is inconsistent with the axially symmetric applied load in the SPS) is due to an insufficient number of crystallites probed during EBSD measurements, although data binning and FWHM of the Gaussian component used during the OD calculation were adapted to meet OD resolution. Consequently, these extra orientations introduced in EBSD data contribute to the decrease of the pole density levels of the {001} pole figure, compared to x-ray refinements.

The uniaxial pressure used during the SPS process results in deformation at high temperature leading to crystallographic slips of individual grains. Slip should occur on specific crystallographic planes along specific crystallographic directions, and the slip planes are generally planes having a high atomic packing density.[18] There are independent slip systems in corundum structure at high temperatures, namely {0001}<11-20>, {1-210}<10-10>, {1-210}<10-11>, {1-102}<01-11> and {10-11}<01-11>.[19] Here, we mainly observe the development of fiber texture in the {0001} direction resulting from basal slip independent of the SPS sintered temperature. At temperatures above 700 °C, the basal slips become predominant [20] with basal critical resolved shear stresses ($\tau_{CB}$) typically lower than 75 MPa. Above 1400 °C, both values of basal slips ($\tau_{CB}$) and prism slips ($\tau_{CP}$) are much lower than 50 MPa, while $\tau_{CB}$ is larger. When the temperature is lower than 700°C, $\tau_{CB}$ becomes larger than $\tau_{CP}$ and prism slip prevails. Thus, the plastic deformation occurring in the low temperature regime leads to the formation of at least one other orientation component with {1-210} and/or {1-102} or {10-11} planes perpendicular to $P_{SPS}$. Since only the {0001} fiber texture is observed, we conclude that under the present experimental SPS conditions {0001}<11-20> is the major activated slip system. Consequently, to accommodate significant axial strain in the hot pressing conditions of SPS, a possible rotation of the grain having the {0001} planes oriented perpendicular to $P_{SPS}$ may occur, and this indeed could be the possible mechanism for $Al_2O_3$ texture development after sliding. This interpretation is supported by the relatively large degree of grain orientation observed. For larger applied pressures, previous texture studies on alumina indicated recrystallization, dislocation slip, grain boundary sliding and anisotropic grain growth effects. [8,17] In these works, the authors successfully achieved {0001} fiber textures with maximum densities as large as 5 times more than the ones observed here, but by using high magnetic fields, higher pressures and temperatures around 1600°C. Nevertheless, our quantitative texture analyses are clearly consistent with thermodynamically favored {0001} slips to withstand uniaxial pressure during the SPS process.[21-23] Misorientation Distribution Functions (MDF), calculated using EBSD data (see Figures 2d and 3d), reache a maximum for {0001} along sample normal direction and increase with temperature probably due to the minimizing grain boundary energy under thermal stress during the SPS process.

The inset of Figure 2b (top one) shows the arrangement of grains having a peculiar misorientation angle. The surrounding grains are probably formed due to the simultaneously uniaxial load and the electric field. The absence of orientation correlation at the grain boundaries with the crystal lattice can be understood by the point to point misorientation profile showing three peaks at grain boundaries (See Inset Fig. 2b). This particular value around 30, 60 and 90 degree high-angle grain boundaries of misorientation angle along the grain boundary can be qualitatively explained by the diffusion of mass transport between the planes of grain boundaries dictated by surface energy, boundary shape and grain size during the SPS consolidation of the starting powders. This means that the value is linked to solid state diffusion which in turn is related to the SPS process and the starting powder (in terms of size and shape). The atoms move more easily from the convex surface on one side of the grain boundary to the concave surface on the other side, rather than in the reverse direction because the chemical potential of the atoms under the convex surface is higher than that for the atoms under the concave surface (for a section across two grains).[8] Also, the driving force for grain growth is the decrease in free energy that accompanies the reduction in the total grain boundary area. Thus, the bimodal grain size distribution (that increases the intensity of small grain size as seen from EBSD data) enhances the packing density leading to almost full density for the 1700°C sample. On the contrary, the average grain size increases but the grain sizes and shapes remain within a fairly narrow range in normal grain growth. Thus, the grain size distribution at a later time is almost similar to the one at an early time. Other factors like the structure and misorientation of the grain boundary could be observed. There is normal unimodal grain size distribution for the 1400 °C indicating normal grain growth. Therefore, at this high sintering temperature (and under 100 Mpa applied pressure), the grain size distribution is controlled. At the end, starting from 8 µm powder, the majority of grain size averages between 17-18 µm (see Figure 2e), indicating a weak distribution. For 1700 °C sample, there is a bimodal grain size distribution indicating abnormal grain growth (see Figure 3e). Surprisingly, the increase in temperature (1700 °C vs. 1400 °C) does not enhance the grain growth. The bimodal grain size distribution could also explain the full densification which can result from an enhanced packing density.

The Young's modulus and hardness measured from nanoindentation experiments, parallel and perpendicular to the direction of uniaxial pressure, are presented in Figure 6. Marked evolution of both Young's modulus and hardness is noticed along the direction of uniaxial pressure. When the sintering temperature is increasing, Young's modulus and hardness increase from 318 to 556 GPa and 15 to 42 GPa, respectively, which is in fact mainly due to an increase in density and fine grained microstructure.[18,24] The substantial increase in anisotropy of mechanical properties is a common consequence of preferred orientations in polycrystalline materials. Grain boundaries can indeed strongly affect nanoindentation measurements giving local values higher than those obtained in the middle of the grains. In consequence, a large spreading of the hardness and Young modulus values appears. Also, higher mean values are obtained on perpendicular samples, which could be related to a higher grain boundary density compared to the one observed for parallel samples. Furthermore the disordered structure of the grain boundary prevents the dislocations from moving in a continuous slip plane, leading to the increase of the required stress depending on the diameter of the grains. [25]

In addition, the degree of elastic anisotropy in polycrystals is governed by two factors, namely the inherent single crystal anisotropy and the distribution of orientations of the constituent grains. For randomly oriented grains, the anisotropic nature of the constituent crystals is averaged out in the bulk making the macroscopic elastic properties isotropic. However, if the constituent crystals exhibit a preferred orientation, the sample will show a macroscopic elastic anisotropy behavior.[26] Therefore, the anisotropic behavior parallel and perpendicular to the uniaxial pressure direction could be associated with the observed texture. In order to estimate the orientation effect on the macroscopic elastic constants, we used the geometric mean approach of Matthies and Humbert,[27] taking the single crystal elastic stiffness tensors from the literature as disposed in the Material Property Open Database [28] and the x-ray refined ODF. Using recent data from Ref. [29], we obtained the following elastic stiffness coefficients: $c_{11}$ and $c_{33}$ = 494 GPa, $C_{44}$=160 GPa, and $c_{12}$, $c_{13}$ and $c_{14}$ = 170 GPa for both samples, suggesting that the resulting macroscopic elastic tensors do not show significant difference between the two samples. Moreover, the texture leads to a nearly isotropic elastic behavior of the sample with $c_{44} = c_{66} = (c_{11}-c_{12})/2$ and $c_{14} = 0$. Such a value is consistent with the fiber character of the orientation and its relatively low texture strength. The literature controversy on the sign of the $c_{14}$ stiffness component, recently assigned positive by Hovis *et al*.,[30] cannot be probed in our simulations. Indeed, all elastic data sets (with positive or negative $c_{14}$) measured up to now would provide similar results, due to the low magnitude of this component compared to the others. However, Bhimasenachar values [31] could not give us a consistent set of elastic stiffnesses through geometric mean. In this case, some of the stiffness Eigenvalues turned out to be negative and without physical meaning, mainly because of an unreasonably large $c_{14}$ component. Also, we could not find in the literature neither $c_{11}$-$c_{33}$ values as large as the ones observed by nanoindentation (typical values are within 1-2% of $c_{33}$), nor $c_{33}$ values larger than 506 GPa (where nanoindentation reaches a Young modulus of 550 GPa). Moreover, our hardness values are 30 % larger than those of Krell and Blank.[32] All these observations suggest that another cause for the increased elastic anisotropy, the large Young moduli and the hardnesses must be proposed. Mao observed strain (Taylor-like) hardening in the Indentation Size Effect (ISE) region with hardness in the 40-47 GPa on (0001)-$Al_2O_3$ single crystals and polycrystals.[33] In the load independent region, the values are calculated as H=27.5±2 GPa for (0001)-$Al_2O_3$ and 30±3 GPa for polycrystals, respectively. They observed that E (0001)-$Al_2O_3$ < polycrystals values for h<100nm, and obtained an averaged E at 466 GPa and 421 GPa for polycrystal and (0001) crystals, respectively, over the first 100 nm. Such values are consistent with previously reported data (cited therein). We have produced 120 nm indenting depths, while Mao et Shen [34] do not observe ISE for depths larger than 60 nm. and then we would not expect as large values as the ones we observe, neither for hardness nor Young moduli. On the other hand, these latter authors do not show evidence of the significant anisotropy of the hardness values between (0001) and (10-12) $Al_2O_3$ single crystal faces, [33-34], while we observe a relatively strong anisotropy for the values measured along and perpendicular to $P_{SPS}$. The anisotropic behavior in Young's modulus and hardness with high value can be attributed to distinct plastic deformation processes under the indenter due to the activation of different slip systems for different indentation surfaces. [35- 38]

When the sintering temperature increases, the misorientation distribution function increases to a misorientation angle of 60 ° due to thepossible ordering of neighbor grains, which are (could be?) also responsible for high hardness along the direction of uniaxial pressure (See Fig. 3(d) and 2(d)). The texture of the 0001 plane along the direction of the uniaxial load could (also?) explain the high hardness and Young's modulus values for alumina sintered at 1700 °C.

**Conclusion**

Spark Plasma Sintering was successfully used to obtain highly textured alumina ceramics. The microstructure and the texture were investigated by using a combination of Electron BackScatter Diffraction and X-rays analysis as a function of the sintering temperature. This analysis reveals that the alumina sintered at 1400 °C is porous with an isotropic grain growth, whereas the 1700 °C sintered sample is less porous with fine grained microstructure but exhibits anisotropic grain growth. The texture computed by combined analysis of XRD and EBSD gives a more intuitive picture on global and local texture. The mechanical behavior of these textured alumina ceramics was found to be slightly superior to that found by previous studies on textured alumina. Thus, we believe that the present study will be important for optimizing the processing conditions in terms of microstructure and texture development, and can be used for potential applications based on textured alumina, such as transparent ceramics.

**Acknowledgements**

We thank E. Guilmeau, J. Lecourt, and X. Larose for their technical help. D.P. and W.P. thank J. Cantwell, R. Klein and S. Bhame for their fruitful discussions. D.P received an Erasmus Mundus PhD fellowship within the IDS FunMat program supported by the European Commission.

**Tables captions:**

Table 1: Refined parameters for the two samples sintered at 1400 and 1700°C. Parentheses indicate r.m.s. standard deviation on the last digit

| Sample | OD min (m.r.d.) | OD max (m.r.d.) | $F^2$ (m.r.d.$^2$) | Cell parameters (Å) | Atomic positions |
|---|---|---|---|---|---|
| 1400°C | 0 | 33.6 | 2.62 | a=4.76168(1) c=13.00019(4) | zAl= 0.35213(1) xO=0.69208(5) |
| 1700°C | 0 | 33.7 | 2.06 | a=4.76321(2) c=13.00365(6) | zAl=0.35207(1) xO=0.69420(6) |

**Figures captions:**

Figure 1: Relative density for a uniaxial pressure of 100 MPa as a function of the dwell time for samples sintered at 1700 °C and 1400 °C. The plot is derived from the SPS data of piston displacement versus time. [8-10] Note that a bimodal distribution is observed for sample grown at 1700 °C as seen from grain size distribution from EBSD data.

Figure 2: (a): Image Quality map (b): Color coded Inverse pole figure (IPF) map showing the grain size for the sample sintered at 1400 °C recorded along the surface perpendicular to uniaxial pressure. The color code corresponding to the crystallographic orientation is given in the stereographic triangle with a small region magnified with the lattice marked with their misorientation angle shown in the top. Wire frames visualize the orientation of the selected crystals. (c): calculated texture index (d): Misorientation Distribution Function (MDF) (e) Grain Size and (f) Misorientation angle Distribution.

Figure 3: (a): Image Quality map (b): Color coded Inverse pole figure (IPF) map showing the grain for the sample sintered at 1700 °C recorded along the surface perpendicular to uniaxial pressure . The color code corresponding to the crystallographic orientation is given in the stereographic triangle with a small region magnified with the lattice marked with their misorientation angle shown in the top. Wire frames visualize the orientation of the selected crystals. (c): calculated texture index  (d): Misorientation Distribution Function (MDF) (e) Grain Size and (f) Misorientation angle Distribution.

Figure 4: Evolution of the 2θ diagrams with the orientation (χ,φ) for the 1400 °C sample (vertical scale). The bottom set is the 864 measured diagrams, while the top set are the fits. For clarity, experimental and fit are indicated.

Figure 5: (a) and (b): {006}, {300} normalized pole figures reconstructed from the ODF, and $P_{SPS}$-direction for the 1400 °C sample (top). (c): Corresponding inverse pole figure of corundum (bottom). The scale used is logarithmic density scale, equal-area density projections.

Figure 6: Hardness and Young's modulus measurement for sample sintered at 1400 °C and 1700 °C MPa for indentation parallel and perpendicular to the direction of uniaxial pressure. (Note: per and para stand for indentation perpendicular and parallel to the direction of uniaxial pressure, respectively)

**References:**

1. J. Morikawa, A. Orie, T. Hashimoto, S. Juodkazis, Opt. Express 18 (2010) 8300.
2. T. Kudrius, G. Slekys, S. Juodkazis, J. Phys. D. Appl. Phys. 43 (2010) 145501.
3. A. Krell, P. Blank, H. Ma, and Thomas Hutzler J. Am. Ceram. Soc., 86 (2003) 12.
4. A. Krell and S. Schädlich, Mater. Sci. Eng. A307 (2001) 172.
5. J.R.G. Evans Journal of the European Ceramic Society 28 (2008) 1421.
6. Z.A. Munir, U. Anselmi-Tamburini, M. Ohyanagi, J. Mater. Sci. 41 (2006) 763.
7. S.W. Wang, L.D. Chen, and T. Hirai, J. Mater. Res., 15 (2000) 982.
8. M.N Rahaman Ceramic Processing and Sintering (2003) Taylor & Francis Editor.
9. J.G. Noudem, D. kenfaui, D. Chateigner and M. Gomina, Scr. Mater. 66 (2012) 258
10. C. Euvananont, N. Jantaping and C. Thanachayanont, Curr. Appl. Phys. 11 (2011) S246.
11. N. Bomshtein, G. Spiridonov and Z. Dashevsky, J. Elec. Mater. 41 (2012) 1546
12. S.I. Wright, M. Nowell and J.F. Binget, Metall. And Mater. Trans. A 38 (2007) 1846.
13. S. Grazulis, D. Chateigner, R.T. Downs, A.F.T. Yokochi, M. Quiros, L. Lutterotti, E. Manakova, J. Butkus, P. Moeck, A. Le Bail, J. Appl. Cryst. 42 (2009) 726.
14. S. Matthies, G. Vinel, K. Helming, Standard Distributions in Texture Analysis. Matthies, Vinel, Helming (Eds). Akademie-Verlag. 1 (1987), 449 pages.
15. W.C. Oliver, G.M. Pharr, J. Mater. Res. 7 (1992) 1564.
16. S.T. Wardle, L.S. Lin, A. Cetel and B.L. Adams, Proc. 52nd Annual Meeting of the Microscopy Society of America, G.W Bailey and A.J. Garratt-Reed, San Francisco Press: San Francisco (1994) 680
17. E. Guilmeau, C. Henrist, T.S. Suzuki, Y. Sakka, D. Chateigner, D.Grossin and B. Ouladdiaf, Mater. Sci. Forum 495-497 (2005) 1395.
18. T. H. Courtney, Mechanical Behaviour of Materials, McGraw-Hill, New York (1990).
19. C. Barry Carter and M. Grant Norton, "Ceramic Materials: Science and Engineering", Springer Science (2007).
20. S.M. Choi and H. Awaji, Sci. and Techn. of Adv. Mater. 6 (2005) 2.
21. H.P. Pinto, R.M. Nieminen and S.D. Elliott, Phys. Rev. B 70 (2004) 125402.
22. A. Marmier, A. Lozovoi, M.W. Finnis, J. Europ. Ceram. Soc.23 (2003) 2729.
23. S. Blonski and S.H. Garofalini, Surf. Sci. 295 (1993) 263.
24. A. Krell, S. Schädlich, Mater. Sci. Eng. A 307 (2001) 172.

25. F. Gao, J. Appl. Phys. 112 (2012) 023506.
26. U. Ramamurty, S. Jana, Y. Kawamura, K. Chattopadhyay, Acta Mater. 53 (2005) 705.
27. S. Matthies and M. Humbert, J. Appl. Cryst. 28, (1995) 254.
28. Giancarlo Pepponi, Saulius Grazulis, Daniel Chateigner: MPOD: a Material Property Open Database linked to structural information. Nuclear Instruments and Methods in Physics Research B 284 (2012) 10.
29. E.H. Kisi, C.J. Howard and J. Zhang, J. App. Cryst. 14 (2011) 216.
30. D.B. Hovis, A. Reddy, and A.H. Heuer, Appl. Phys. Lett. 88 (2006) 131910.
31. J. Bhimasenachar, Proc. National Inst. Sci. Ind. 16 (1950) 242.
32. A. Krell and J. Blank, J. Amer. Ceram. Soc. 78 (1995) 1119.
33. W.G. Mao, Y.G. Shen, and C. Lu Scripta Materialia 65 (2011) 127.
34. W. Mao and Y. Shen, J. Am. Ceram. Soc 95 (2012) 3605.
35. W.G. Mao, Y.G. Shen, and C. Lu, J. Europ. Cerm. Soc. 31 (2011) 1865.
36. M. Kaji, M.E. Stevenson and R.C. Bradt, J. Am. Ceram. Soc 85 (2002) 415
37. K. Nishimura, R.K. Kalia, A. Nakano and P. Vashishta, Appl. Phys. Lett. 92 (2008) 161904.
38. A.H. Heuer, N.J. Tighe, R.M. Cannon, J. Am. Ceram. Soc. 63 (1980) 1.

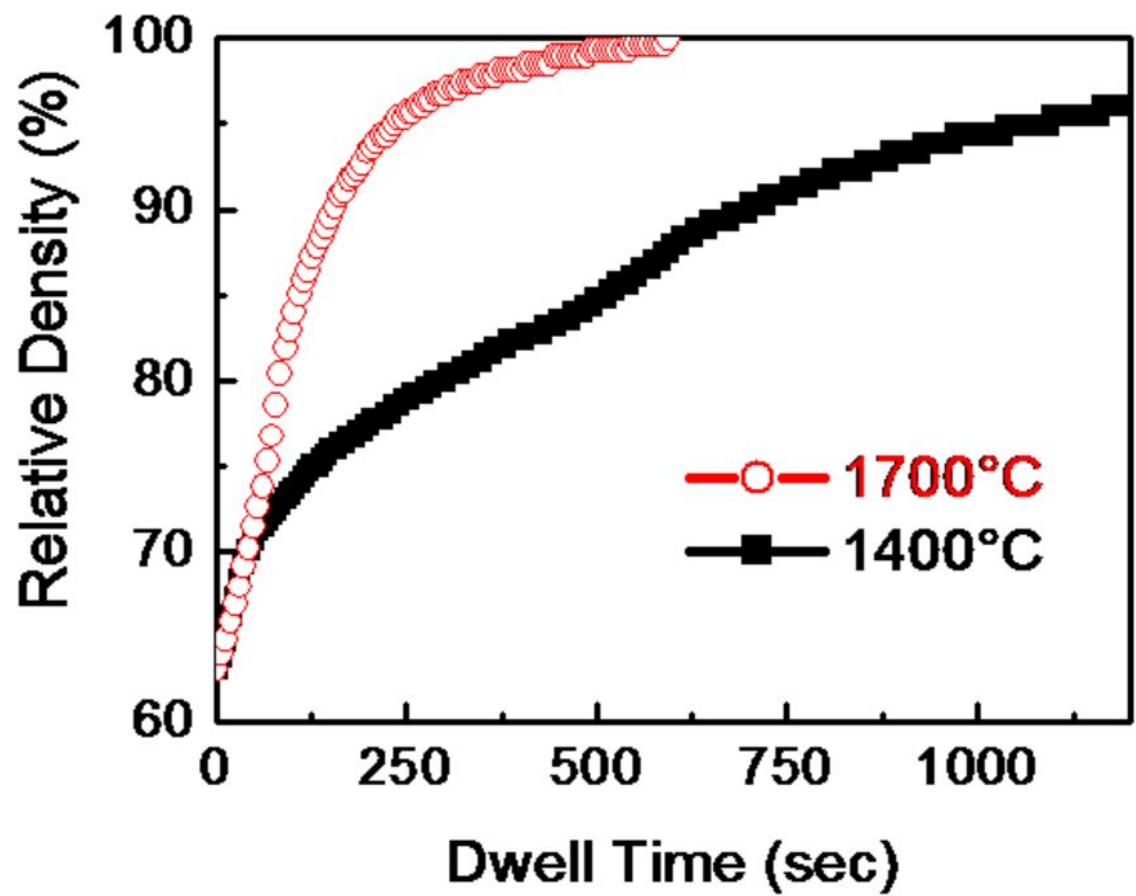

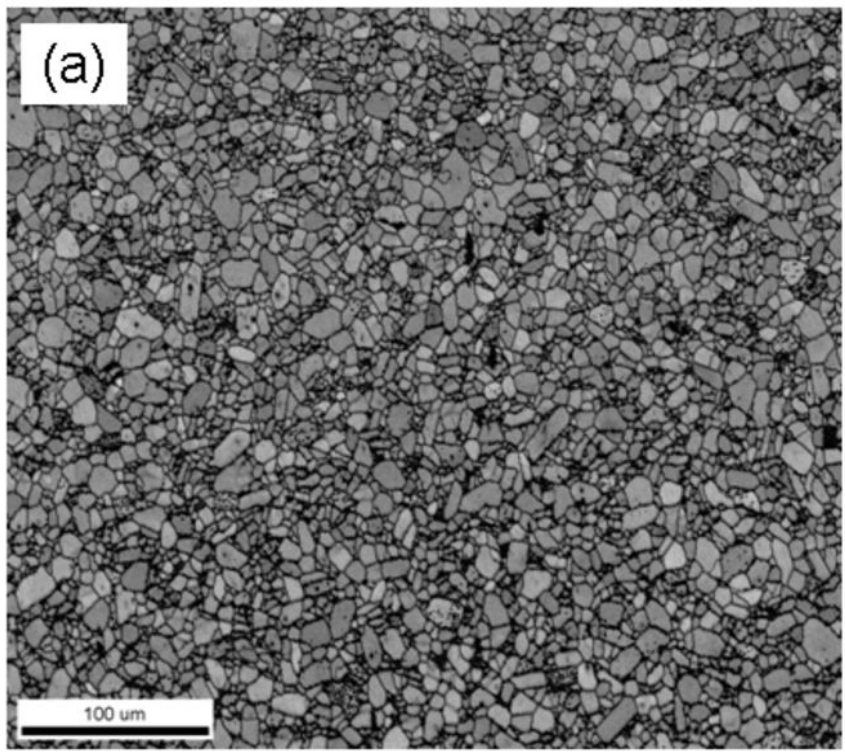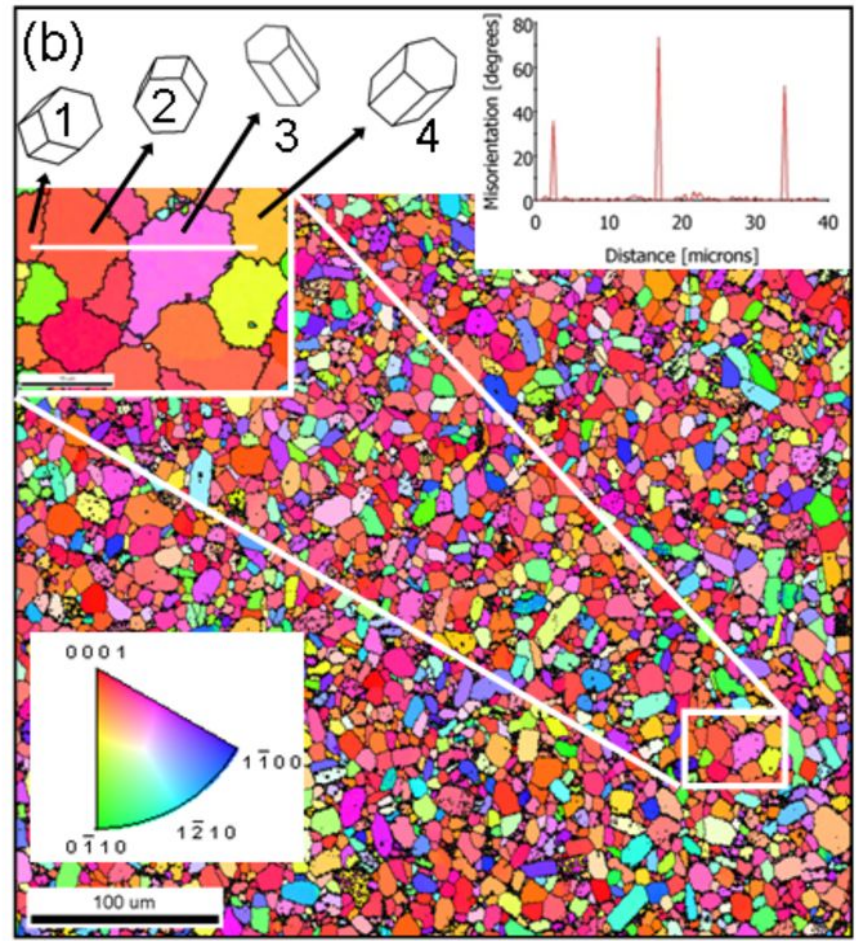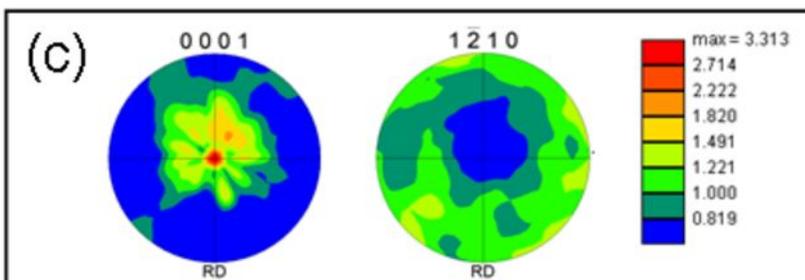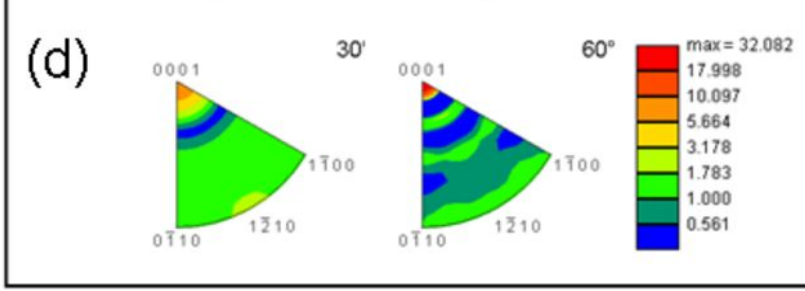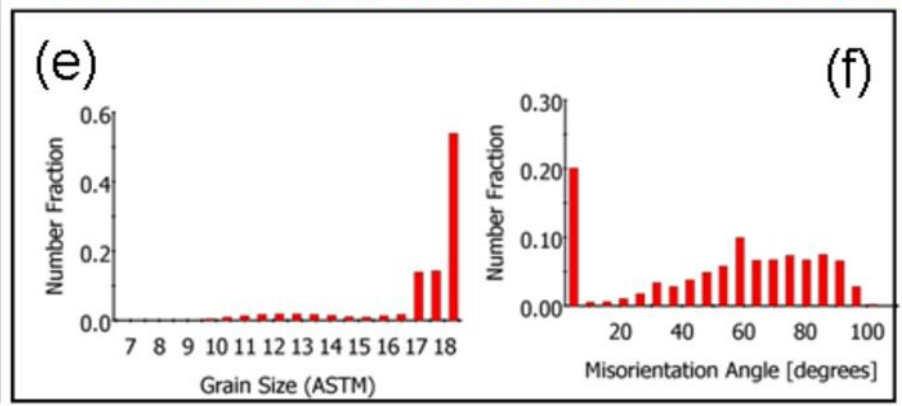

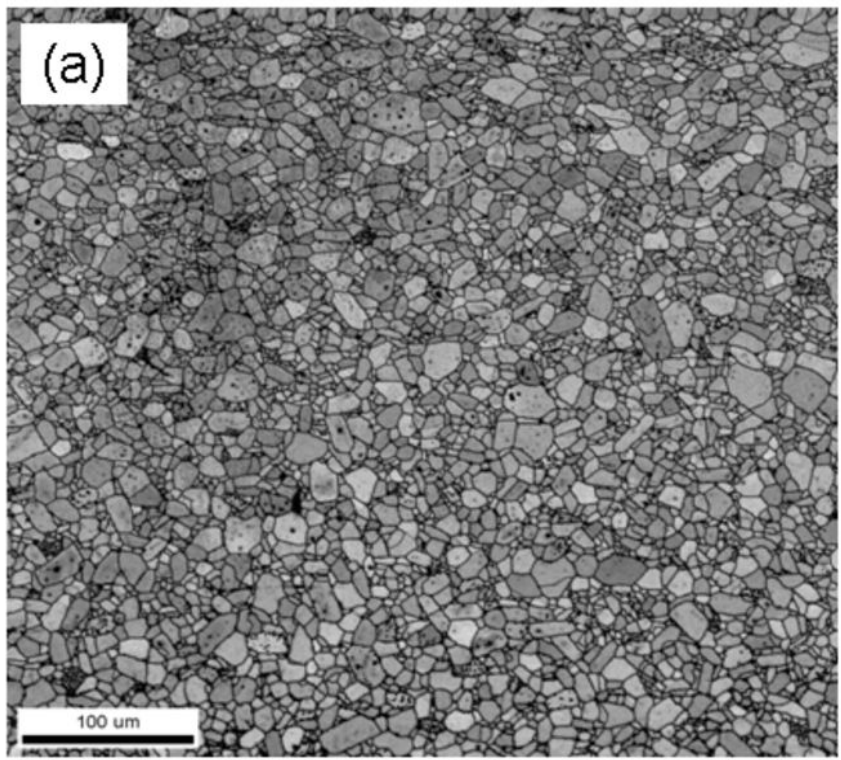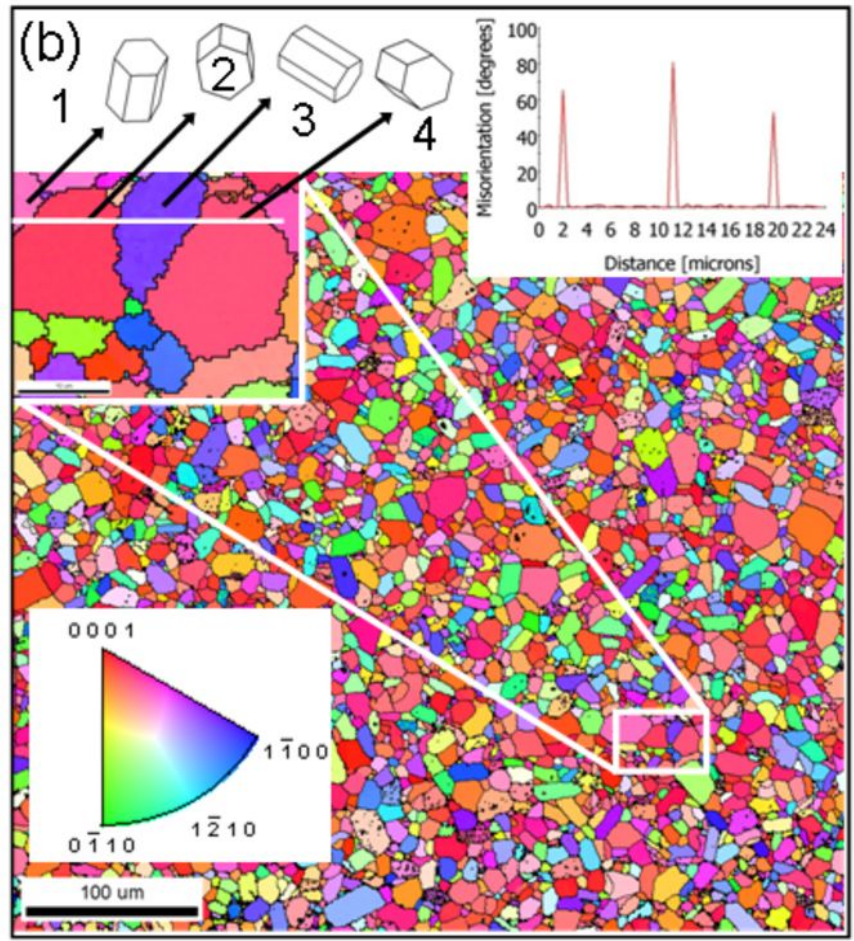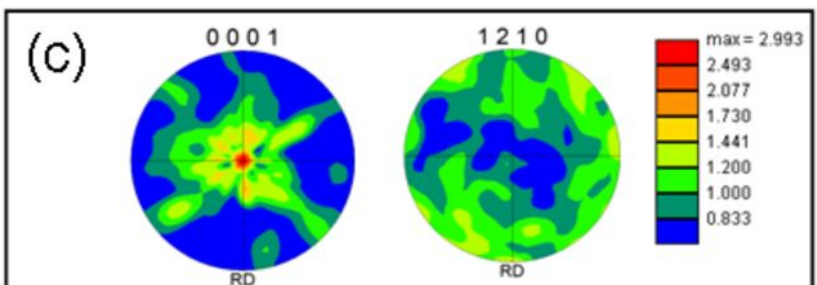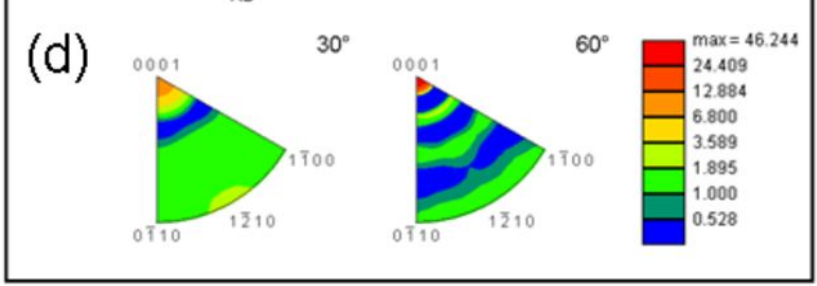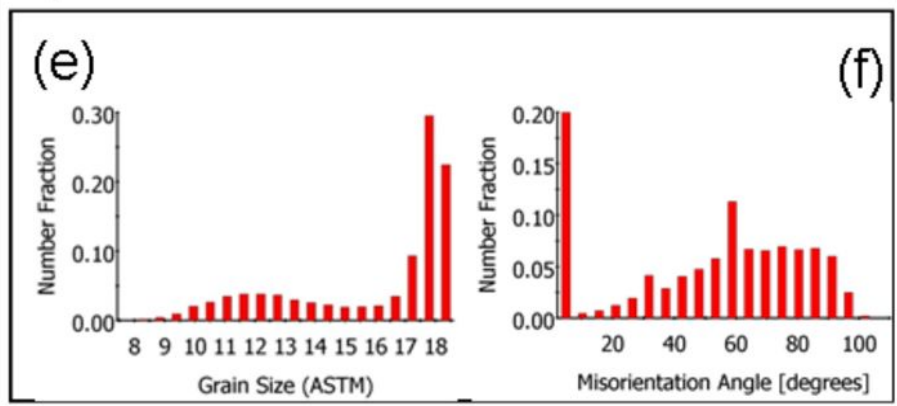

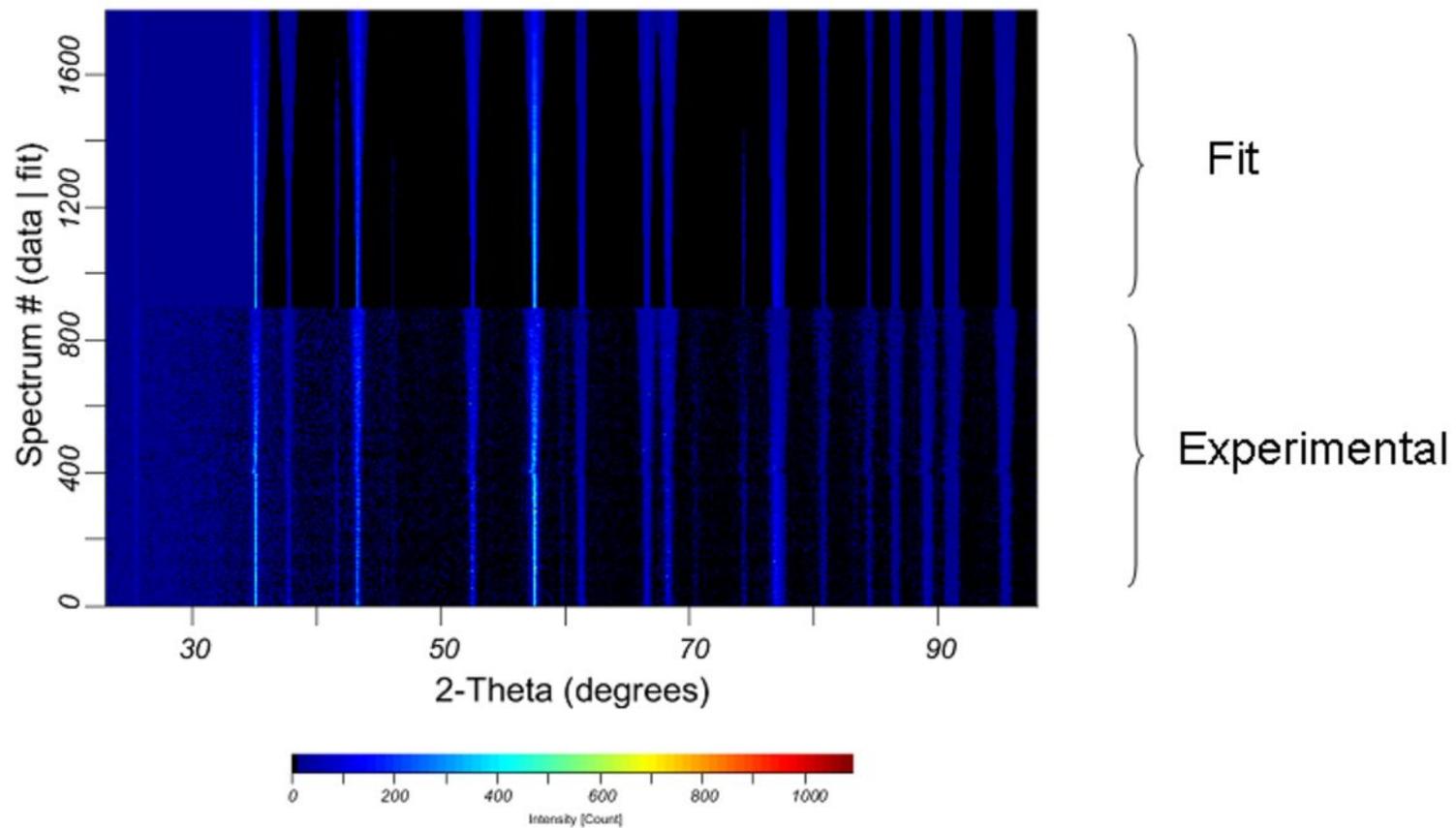

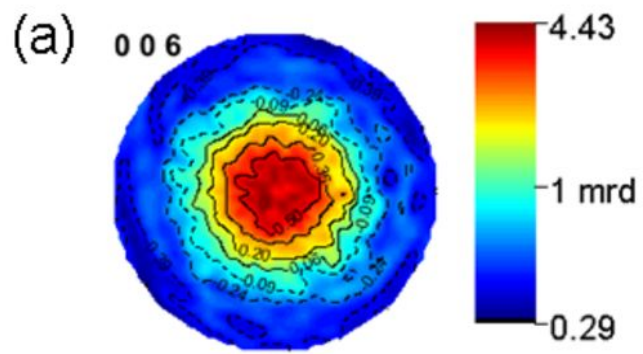

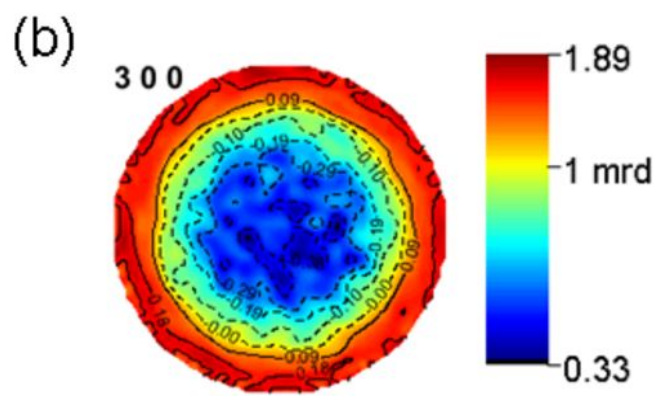

Rw (%) = 18.143974
Rb (%) = 18.311705

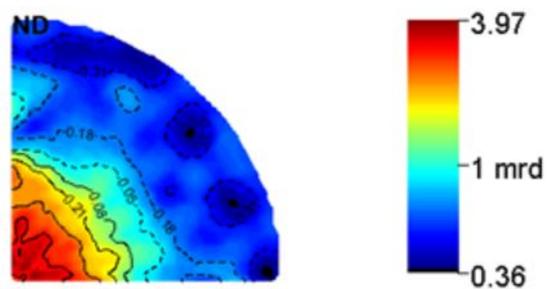

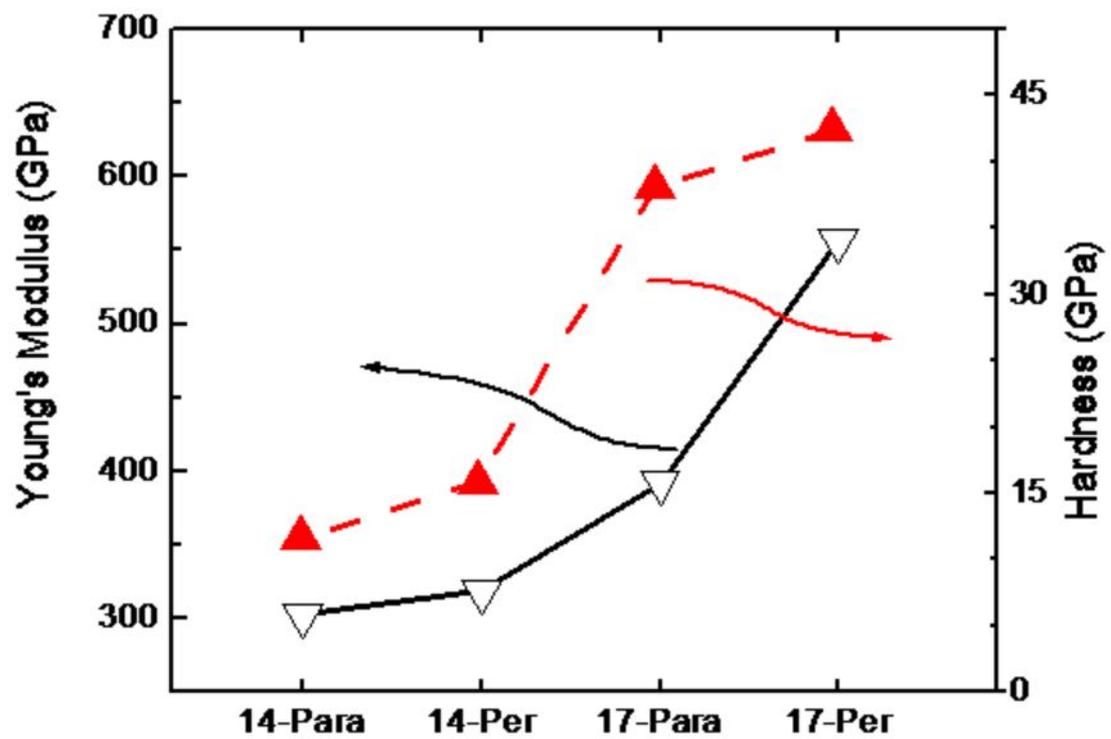